\documentclass[12pt,a4paper,preprintnumbers,superscriptaddress,showkeys,]{revtex4-1} 

\usepackage[fleqn]{amsmath}
\usepackage{amssymb}
\usepackage{bm}
\usepackage{graphicx} 
\usepackage{graphics}
\graphicspath{{Figures/}}
\usepackage{subfigure}
\usepackage[pdfstartview=FitH,
            CJKbookmarks=true,
            bookmarksnumbered=true,
            bookmarksopen=true,
            colorlinks=true,
            pdfborder=001,
            linkcolor=blue,
            citecolor=blue,
            urlcolor=blue
            ]{hyperref}
\RequirePackage{color}
\renewcommand{\thetable}{\arabic{table}} %
\renewcommand{\figurename}{Fig.} %
\renewcommand{\tablename}{Table}  %

\begin{document}


\title{Compression-induced stiffness in the buckling of a one fiber composite}

\author{R. Lagrange}
\affiliation{Massachusetts Institute of Technology, Department of Applied Mathematics,  Cambridge, MA 02139-4307, USA}

\date{\today}

\begin{abstract}
\noindent \textbf{Abstract} We study the buckling of a one fiber
composite whose matrix stiffness is slightly dependent on the
compressive force. We show that the equilibrium curves of the system
exhibit a limit load when the induced stiffness parameter gets
bigger than a threshold. This limit load increases when increasing
the stiffness parameter and is related to a possible localized path
in the post-buckling domain. Such a change in the maximum load may
be very desirable from a structural stand point.
\end{abstract}

\keywords{Buckling, Fiber composite, Initial curvature,
Compression-induced stiffness, Limit point}

\maketitle

\section{Introduction}

Important engineering applications such as railway tracks lying on a
soil base, thin metal strips attached to a softer substrate or
structures floating on fluids require accurate modeling of a layer
bonded to a substrate-foundation. \cite{Shield94} and more recently
\cite{Bigoni2008,Sun2012} have shown that a beam theory model for
the layer and a Winkler-type springs model for the foundation is
accurate enough to correctly describe the layer-substrate system.
The restoring force provided by the springs may depend linearly or
nonlinearly on the local displacement. Many analytical or numerical
analysis considered the mechanical response of a straight elastica
attached to a linear foundation, see e.g.
\cite{Lee96,Kounadis2006,Suhir2012}. In addition, the mechanical
behavior of a beam, initially straight or with an initial curvature,
lying on a nonlinear elastic foundation, have been the subject of
many studies, see e.g. \cite{Potier15,Hunt93,
HuntBlack,Wadee97,Wadee18,Whiting17,Netto99,
Zhang2005,Jang2011,Lagrange2012}. An important point to notice is
that the equilibrium curves may exhibit in this case a limit point,
i.e. a maximum load, and a bifurcation point in the post-buckling
path, related to a localized mode.

In the present paper, we use a beam on foundation model to analyze
the static equilibrium of a one fiber composite suffering a
compressive stress. The introduction of an initial curvature in the
line of the beam models the case of a slightly misaligned fiber in
the direction of compression. A restoring force, function of the
compressive load, takes into account the dependence of the matrix
stiffness with the overall compression. Such a dependence has
received very little attention since the work of \cite{Waas90} who
studied the initial post-buckling of a curved fiber on a cubic
foundation. Using an asymptotic expansion of the equilibrium
equation about the critical load, he showed that the
compression-induced stiffness affects the buckling and post-buckling
behavior of the fiber.

The present paper aims to extend these results to the case of a
bi-linear foundation whose stiffness is affinely dependent on the
compressive force. Of particular interest, we will focus on the
existence and the evolution of a limit point (i.e. saddle-node) in
the bifurcation diagram of the system.

\section{Problem formulation}\label{Section2}
\begin{figure}
\begin{center}
\includegraphics[width=0.5\textwidth]{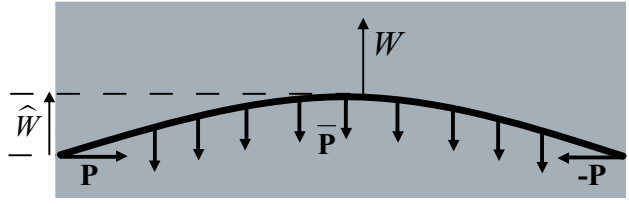}
\end{center}
\caption{Sketch of a one fiber composite undergoing a compressive
load ${\bf{P}}$. The fiber has an initial imperfection $\widehat{W}$
and its lateral displacement is $W$. The matrix restoring force per
unit length ${\bf{\overline{P}}}$ is a function of the compressive
load and the lateral deflection.}\label{Limit_point}
\end{figure}

We consider a fiber of length $L$, bending stiffness $EI$, subjects
to a compressive force $P$ at both ends. The fiber has an initial
imperfection shape $\widehat{W}=\widehat{\tau}\sin \left({\pi
}X/L\right)$, with $\widehat{\tau}$ the amplitude and $X$ the
longitudinal coordinate.

The fiber and matrix are assumed to be well bonded at their
interface and remain that way during deformation. Thus, interfacial
slip, fiber/matrix debonding or matrix micro-cracking is not
considered. The supporting matrix is modeled as a Winkler type
foundation providing a lateral restoring force per unit length
$\overline{P}$, function of the compressive load $P$ and the lateral
deflection $W$

\begin{eqnarray}\label{LoiPorpor2}
\overline{P}\left( W \right)= \left(K+\Sigma P\right)\left\{
\begin{array}{l}
  -  W\hspace{0.9cm}{\rm{ if}}\hspace{0.1cm}\left| W \right| < \Gamma , \\
  -  \Gamma \hspace{1.1cm}{\rm{ if }}\hspace{0.38cm}W > \Gamma {\rm{ }}, \\
   \Gamma \hspace{1.40cm}{\rm{   if }}\hspace{0.38cm}W <  - \Gamma.
 \end{array} \right.
\end{eqnarray}
In the above expression, $K+\Sigma P$ is the stiffness of the
supporting matrix and $\Gamma$ its mobilization (also named the
yield point). Compression-induced hardening (resp. softening)
corresponds to $\Sigma>0$ (resp. $\Sigma<0$).

We note $L_{c} = \left( {{{EI}}/{K}} \right)^{\frac{1}{4}}$ a
characteristic length of the problem, and define the non-dimensional
quantities
\begin{eqnarray}\label{GrandeursAdimBis2}
l &=& \frac{{ L }}{{L_{c} }},\quad x= \frac{{ X }}{{L_{c} }}, \quad
w = \frac{{W}}{{\Gamma}}, \quad \widehat{w} =
\frac{{\widehat{W}}}{{\Gamma}}, \quad
\tau=\frac{\widehat{\tau}}{\Gamma}, \quad \sigma=\Sigma {L_c}^2,
\quad\lambda = \frac{P {L_{c}}^2}{{EI }},
\end{eqnarray}
as respectively the dimensionless fiber length, longitudinal
coordinate, lateral deflection, imperfection shape, imperfection
size, stiffness parameter and compressive load. The restoring force
may be recast in a convenient computational form as

\begin{eqnarray}\label{Functionp}
\overline{p}\left( w \right) =  - w -\left( {{\mathop{\rm sgn}}
\left( w \right) - w} \right){\rm{H}}\left( {\left| w \right| - 1}
\right),
\end{eqnarray}
where $\rm{sgn}$ denotes the sign function and $\rm{H}$ is the
Heaviside function, defined as ${\rm{H}}\left( \left| w \right| -
1\right)=0$ if $\left| w \right|<1$ and $1$ if $\left| w \right|>1$.

Assuming that $\lambda$ and $\overline{p}$ are conservative, the
deflection equation is derived using an energy formulation. Strains
are assumed to be small compared to unity. The centroidal line is
inextensible and cross sections remain normal to this line (i.e.
Euler-Bernoulli assumption). The imperfection is assumed to be small
so that nonlinearities in $\widehat{w}$ or $\widehat{w}'$ are
dropped in the formulation of the potential energy. Under these
assumptions, the potential energy with low-order geometrically
nonlinear terms is \cite[see][]{Potier15}

\begin{eqnarray}\label{EnergiePotentielle}
V = \int\limits_0^l {\left[ {\frac{1}{2}{w^{''}} ^2  -
\lambda\left({\frac{1}{2}{w^{'}}^2 +{{\widehat{w}}^{'}}{{w}^{'}}
}\right) - \int\limits_0^w {\overline p \left( t \right){\rm{dt}}} }
\right]{{\rm{{dx}}}}},
\end{eqnarray}
where a prime means $d/dx$. In (\ref{EnergiePotentielle}),
$1/2{w^{''}}^2$ is the bending energy, $\lambda\left({{1/2}{w^{'}}^2
+{{\widehat{w}}^{'}}{{w}^{'}} }\right)$ the work done by the
compressive force $\lambda$ and $\int\limits_0^w {\overline p \left(
t \right){\rm{dt}}}$ the elastic foundation energy. Note that
equation (\ref{EnergiePotentielle}) is written in term of the
displacement field $w$ measured from the initial configuration, but
an equivalent formulation may be derived \cite[see][]{Wadee18}
introducing the vertical coordinate $y=\widehat{w}+w$ of the
centroidal line. Here, we use the formulation
\ref{EnergiePotentielle} because the restoring force $\overline{p}$
is more easily expressed in term of the displacement field $w$ than
in term of the vertical position $y$.

Equilibrium states are critical values of $V$. Assuming a simply
supported fiber (i.e. kinematic boundary conditions are $w\left( 0
\right)=0$ and $w\left( l \right)=0$), variations of
(\ref{EnergiePotentielle}) for an arbitrary kinematically admissible
virtual displacement $\delta w$ leads to the Euler-Lagrange equation
(also named the stationary Swift-Hohenberg equation)

\begin{eqnarray}\label{EqDiff}
{w^{''''}  + \lambda \left( {w^{''}  + \widehat{w} ^{''} } \right) -
\left(1+\sigma\lambda\right) \overline{p}\left( w \right)}=0,
\end{eqnarray}
along with static boundary conditions $w''\left( 0 \right)=
w''\left( l \right)=0$.

This equation is nonlinear because of the restoring force and is
highly sensitive to the parameters. Consequently it is illusory to
depict the behavior of the system over a wide range of parameters.
The imperfection size $\tau=\widehat{\tau}/\Gamma$ is kept of order
$O(1)$ so that $\widehat{\tau}$ and the matrix mobilization $\Gamma$
are of the same order, which is typically less than the millimeter.
Bigger values for $\tau$ would correspond to intentionally
misaligned fiber in the direction of compression, what is out of
consideration in the present article. The stiffness parameter
$\sigma$ is kept of order $o(1)$, based on the idea that the
compression-induced stiffness $\left| \Sigma \right| P$ remains
small compared to the linear stiffness $K$, for any compressive load
$P$.

\section{Results}\label{SectionDiscussion}
\begin{figure}
\begin{center}
\includegraphics[width=1\textwidth]{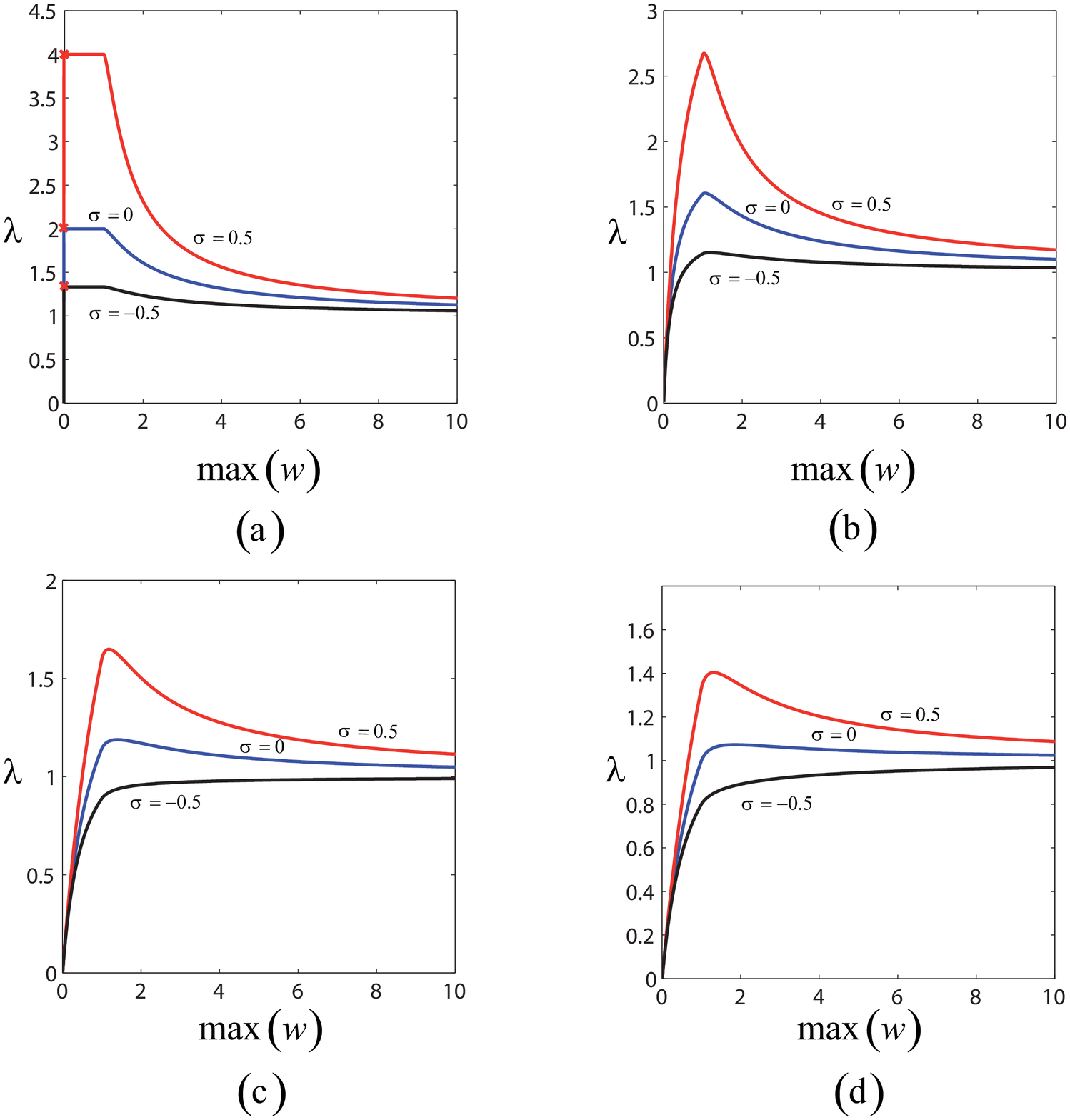}
\end{center}
\caption{Equilibrium paths of a one fiber composite plotted for (a)
$\tau=0$, (b) $\tau=0.25$, (c) $\tau=0.75$ and (d) $\tau=1$. A red
cross indicates the critical buckling load
$\lambda_c=2/\left(1-\sigma\right)$. The fiber length is
$l=\pi$.}\label{CourbesEquilibre}
\end{figure}

\begin{figure}
\begin{center}
\includegraphics[width=1\textwidth]{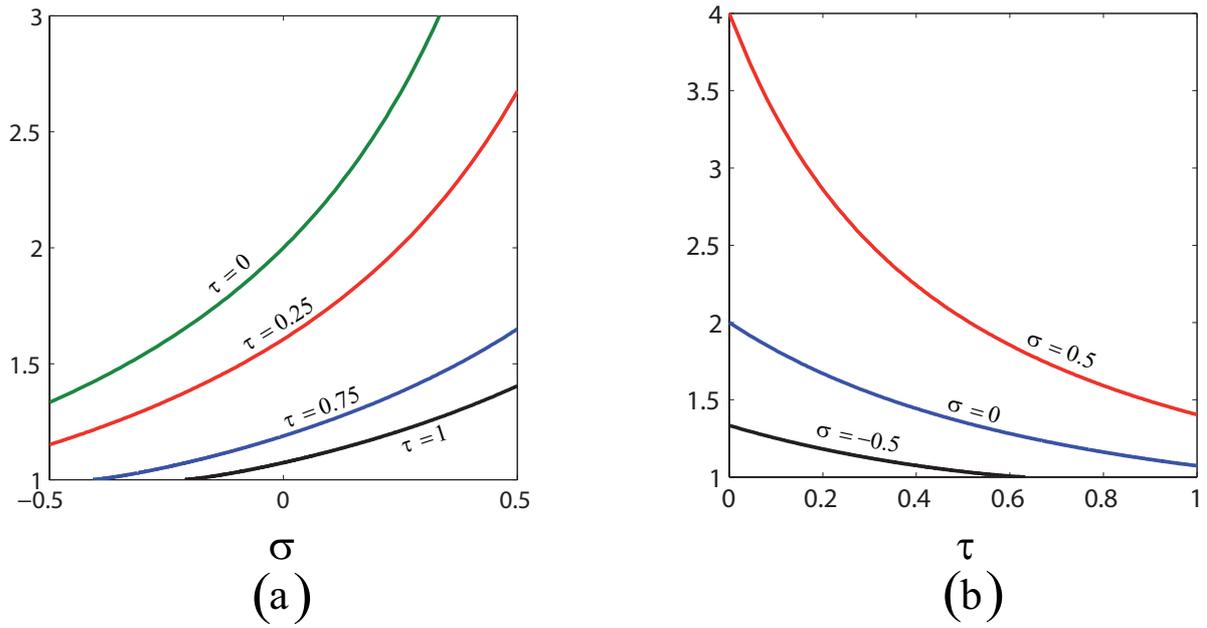}
\end{center}
\caption{Limit load $\lambda_m$ of a fiber composite vs. (a) the
matrix stiffness parameter $\sigma$, (b) the initial imperfection
size $\tau$. The fiber length is $l=\pi$. }\label{Limit_point}
\end{figure}

\begin{figure}\label{Diagram}
\begin{center}
\includegraphics[width=2.8in]{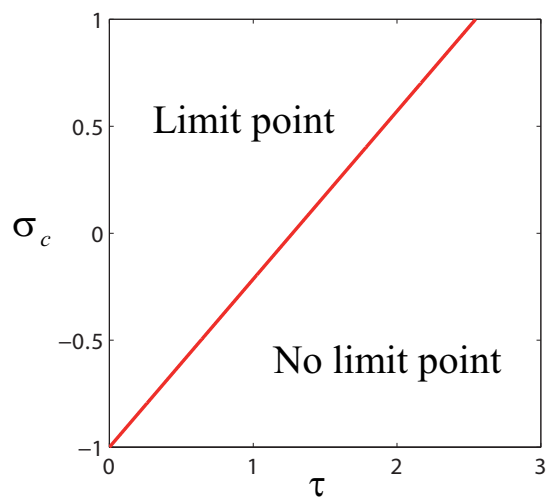}
\end{center}
\caption{Critical matrix stiffness parameter $\sigma_c$ vs.
imperfection size $\tau$, leading to the existence of a limit point
in the equilibrium curves of a one fiber composite. The fiber length
is $l=\pi$.}\label{Sigmac}
\end{figure}

\begin{figure}
\begin{center}
\includegraphics[width=1\textwidth]{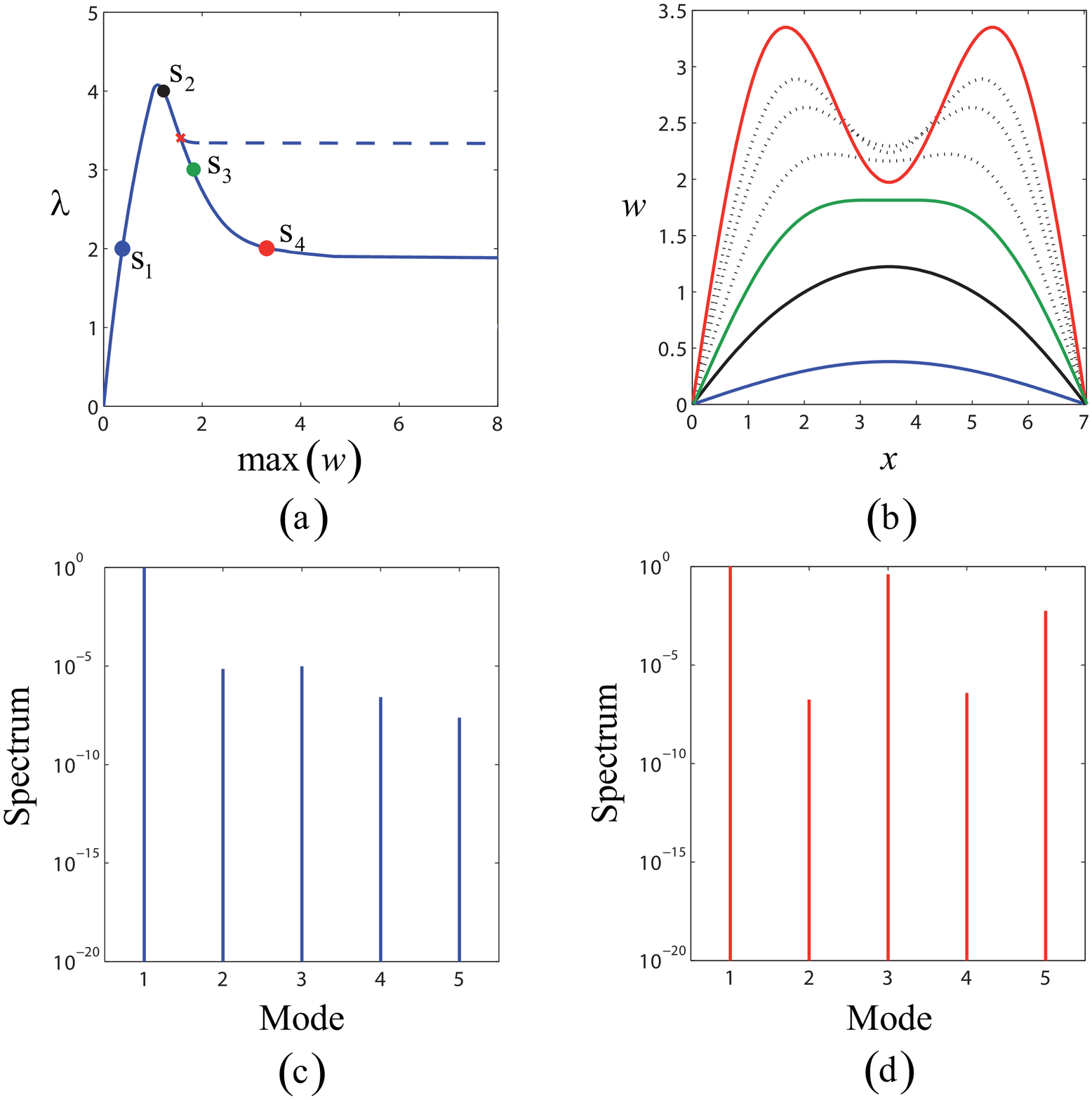}
\end{center}
\caption{(a) Equilibrium path for $\sigma=0.1$, $\tau=0.8$, $l=7$.
${\rm{s}}_1$, ${\rm{s}}_2$, ${\rm{s}}_3$ and ${\rm{s}}_4$ are some
equilibrium states for $\lambda=2$, $\lambda=4$, $\lambda=3$ and
$\lambda=2$. Red cross indicates a point of bifurcation. (b)
Deflective patterns of the equilibrium states ${\rm{s}}_1$ (blue curve),
${\rm{s}}_2$ (black curve), ${\rm{s}}_3$ (green curve) and
${\rm{s}}_4$ (red curve). Dotted curves show post-buckling patterns
for equilibrium states between ${\rm{s}}_3$ and ${\rm{s}}_4$. (c)
Spectrum of the equilibrium state ${\rm{s}}_1$. (d) Spectrum of the
equilibrium state ${\rm{s}}_4$. }\label{Patterns}
\end{figure}

Equation (\ref{EqDiff}) is solved using the MATLAB's routine
{\it{bvp4c}} \cite[this boundary value solver uses a finite
difference method that executes a collocation formula, see][for more
details]{Matlab}. Equilibrium paths are traced out in the plane
$\left({\rm{max}}\left(w\right), \lambda\right)$ by gradually
incrementing $\lambda$.

For a perfect fiber ($\tau=\widehat{w}=0$), $w=0$ satisfies
(\ref{EqDiff}), for any $\sigma$. This solution is stable up to a
point of bifurcation from which a new path emanates. This path is
horizontal in the linear domain of the restoring force and decreases
to an horizontal asymptote in the plastic domain. The point of
bifurcation is classically determined by linearizing the equilibrium
equation (\ref{EqDiff}) about $w=0$ and looking for sinusoidal
solutions $\sin \left({n\pi }x/l\right)$, $n$ being an integer. It
is found that sinusoidal solutions may arise for $\lambda_n=
{{\left[ {1 + \left( {{{n\pi } \mathord{\left/
 {\vphantom {{n\pi } l}} \right.
 \kern-\nulldelimiterspace} l}} \right)^4 } \right]} \mathord{\left/
 {\vphantom {{\left[ {1 + \left( {{{n\pi } \mathord{\left/
 {\vphantom {{n\pi } l}} \right.
 \kern-\nulldelimiterspace} l}} \right)^4 } \right]} {\left[ {\left( {{{n\pi } \mathord{\left/
 {\vphantom {{n\pi } l}} \right.
 \kern-\nulldelimiterspace} l}} \right)^2  - \sigma } \right]}}} \right.
 \kern-\nulldelimiterspace} {\left[ {\left( {{{n\pi } \mathord{\left/
 {\vphantom {{n\pi } l}} \right.
 \kern-\nulldelimiterspace} l}} \right)^2  - \sigma } \right]}}$,
 leading to a critical buckling load $\lambda_c=\lambda_1=2/\left(1-\sigma\right)$ for $l=\pi$ and
 $\sigma<3/5$.

Results for an imperfect fiber are plotted in Fig.
\ref{CourbesEquilibre}(b, c, d), showing two types of equilibrium
paths.

For small $\sigma$ (resp. high $\tau$), the equilibrium paths are
increasing and they tend to an horizontal asymptote when
${\rm{max}}\left(w\right)\rightarrow\infty$.

For high $\sigma$ (resp. small $\tau$), the equilibrium paths are
increasing at first (pre-buckling domain), then they hit a maximum
(limit point), and eventually they decrease (post-buckling domain)
to the previous cited asymptote when
${\rm{max}}\left(w\right)\rightarrow\infty$.

Moreover, when decreasing (resp. increasing) the stiffness parameter
$\sigma$ (resp. the imperfection size $\tau$), the equilibrium paths
flatten out, leading to a progressive drop in the maximum force
$\lambda_m$ and a gradual increase in the maximum displacement. The
variations for $\lambda_m$ are confirmed in Fig.
\ref{Limit_point}(a, b). For $\sigma$ (resp. $\tau$) smaller (resp.
bigger) than a critical value $\sigma_c$ (resp. $\tau_{c}$) there is
no more limit point in the bifurcation diagram. Iterating over
$\sigma$ and $\tau$ the procedure for plotting an equilibrium path,
tracking the limit point at each step, results in the $\sigma_c$ vs.
$\tau$ plot shown in Fig. \ref{Sigmac}. From this figure it appears
an affine dependence of $\sigma_c$ on $\tau$

\begin{eqnarray}\label{SigmaCrit}
\sigma _{\rm{c}}  = {\rm{f}}\left( {l } \right)\tau  - {\mathop{\rm
g}\nolimits} \left( {l } \right),
\end{eqnarray}
with ${\rm{f}}$ and ${\rm{g}}$ two scaling functions of $l$.

Finally, deflective patterns are shown in Fig. \ref{Patterns}, along
with their Fourier spectrums. It is observed that a pre-buckling
state has a fundamental harmonic along the first buckling mode,
other harmonics being negligible. Consequently, the deflective
pattern of a pre-buckling state is an amplification of the initial
curvature. This feature holds in the initial post-buckling domain,
up to a subcritical point of bifurcation from which a localized path
emanates (path containing ${\rm{s}}_3$ and ${\rm{s}}_4$ in Fig.
\ref{Patterns}(a)). An equilibrium state lying on this path exhibits
a deflection that is no longer an amplification of the initial
curvature, but a combination of various harmonics whose spectrum
amplitudes are increasing in the far post-buckling domain.

In terms of stability, near a subcritical bifurcation, periodic
responses require more energy to trigger and hence will not appear
in physical test. On the contrary, the stability of the localized
responses depends on the loading condition. As stated in
\cite{Peletier}, localized modes are generally unstable under dead
loading (i.e. experiments in which the compressive force is the
controlled parameter) and stable under rigid loading (i.e.
experiments in which the displacement is the controlled parameter).

\section{Conclusion}

This paper considers the matrix stiffness effects on the buckling
behavior of an imperfect fiber in a material composite. The
imperfection has been introduced as an initial curvature and the
matrix stiffness taken as compressive dependent.

The compression-induced stiffness and the imperfection size play
antagonistic roles in the buckling response of the fiber. Hardening
(respectively softening) leads to an increase (respectively
decrease) of the limit load. On the contrary, an increase
(respectively decrease) of the imperfection size leads to a decrease
(respectively increase) of the limit load. Such a limit load does
not exist any more for a critical stiffness $\sigma_c$ dependent on
the imperfection size and the fiber length. Note that these features
are in agreement with the theoretical predictions of \cite{Waas90},
carried out for a cubic foundation.

Finally, the gradual decrease of the limit point with the
compression-induced stiffness could explain the progressive
transition to final failure that occurs in a composite material
\cite[see][]{Hahn}.

The present paper has to be considered as a preliminary study,
focusing mainly on the existence and the variations of a limit point
with the foundation stiffness parameter and the initial curvature.
Future works should determine the influence of these two parameters
on the post-buckling domain, in particular on the existence and the
behavior of a bifurcation point leading to a localization, as
depicted in Fig. \ref{Patterns}. Taking into account the
higher-order geometrically nonlinearities in the potential energy,
we aim to explore the far post-buckling domain through the
path-following and bifurcation analysis software MANLAB
\cite[see][]{Cochelin2007}.
\\

The author acknowledges Dr. Alban Sauret for his insightful comments
on this paper.


\bibliographystyle{aipnum4-1}
\bibliography{Biblio}


\end{document}